\documentclass[namedreferences]{SolarPhysics}
\usepackage[optionalrh]{spr-sola-addons} % For Solar Physics 
\usepackage{graphicx}        % For eps figures, newer & more powerfull
\usepackage{color}           % For color text: \color command
\usepackage{url}             % For breaking URLs easily trough lines
\begin{document}

\begin{article}

\begin{opening}

\title{Solar Flux Emergence Simulations}

\author{R.~F.~\surname{Stein}$^{1}$\sep
        A.~\surname{Lagerfj{\"a}rd}$^{2}$\sep
        {\AA}.~\surname{Nordlund}$^{2}$\sep      
        D.~\surname{Georgobiani}$^{1}$      
       }
\runningauthor{Stein et al.}
\runningtitle{Flux Emergence Simulations}

   \institute{$^{1}$ Michigan State University, East Lansing, MI, USA
                     email: \url{stein@pa.msu.edu, dali@pa.msu.edu}\\ 
              $^{2}$ Astronomical Observatory/Niels Bohr Institute, Juliane Maries Vej 30, DK-2100 Copenhagen, DK
                     email: \url{andersl@astro.ku.dk, aake@astro.ku.dk} \\
             }

\begin{abstract}
We simulate the rise through the upper convection zone and emergence
through the solar surface of initially uniform, untwisted, horizontal 
magnetic flux with the same entropy as the non-magnetic plasma that is
advected into a domain 48 Mm wide from from 20 Mm deep.  The magnetic 
field is advected upward by the diverging upflows and pulled down in the
downdrafts, which produces a hierarchy of loop like structures of
increasingly smaller scale as the surface is approached.  
There are significant differences between the behavior of fields of 10 kG 
and 20 or 40 kG strength at 20 Mm depth.  The 10 kG fields have little
effect on the convective flows and show little magnetic buoyancy effects, 
reaching the surface in the typical fluid rise time from 20 Mm depth of 
32 hours.  20 and 40 kG fields significantly modify the convective flows,
leading to long thin cells of ascending fluid aligned with the magnetic 
field and their magnetic buoyancy makes them rise to the surface faster 
than the fluid rise time.  The 20 kG field produces a large scale magnetic 
loop that as it emerges through the surface leads to the formation of a 
bipolar pore-like structure.
\end{abstract}

\end{opening}
%-------------------------------------------------
% Outline:
% Flux Emergence
%	rise times
%	buoyancy
%	10kG case
%	20kG case
%	spreads to entire region
%	collects to form pores
%	shallow field structure
%Depth dependence
%	rho^(0.33)
%Loop Structures
%	large scale
%	small scale near surface

\section{Introduction}\label{Introduction} 

We have initiated simulations of the rise of magnetic flux through the
near surface layers of the solar convection zone.  Numerous studies
exist of the rise of a thin magnetic flux tube through the bulk of the
convection zone, e.g.
\cite{Moreno-Insertis86,Choudhuri87,Fan94,Moreno-Insertis94,Caligari95}.
These breakdown near the bottom of our simulation because the magnetic
pressure exceeds the gas pressure and the tubes explode.  Others have
studied the rise of coherent, twisted, non-thin flux tubes through
model convection zones, e.g.  \cite{Dorch03,Abbett04,Fan09}.  The rise
of a coherent, twisted flux tube through the shallow layers of
realistic solar convection zones has been studied by
\cite{Cheung08,Cheung07,Cheung06,Martinez-Sykora09,Martinez-Sykora08}.
We are undertaking a complimentary study -- the rise and evolution of
untwisted, horizontal flux with the same entropy as the unmagnetized 
ascending plasma, carried by upflows from a depth of 20 Mm.
Here we present some preliminary results from these simulations.

\section{Simulation}\label{Simulation} 

\begin{figure}    %%%%%%%%%%%%%%%%%% FIGURE 1 
%  \centerline{\includegraphics[width=0.47\textwidth]{atm_P_T_r_S_ym.pdf}
%  \includegraphics[width=0.62\textwidth,clip=]{atm_ion_gam_ym.pdf}}
   \centerline{\includegraphics[width=0.47\textwidth]{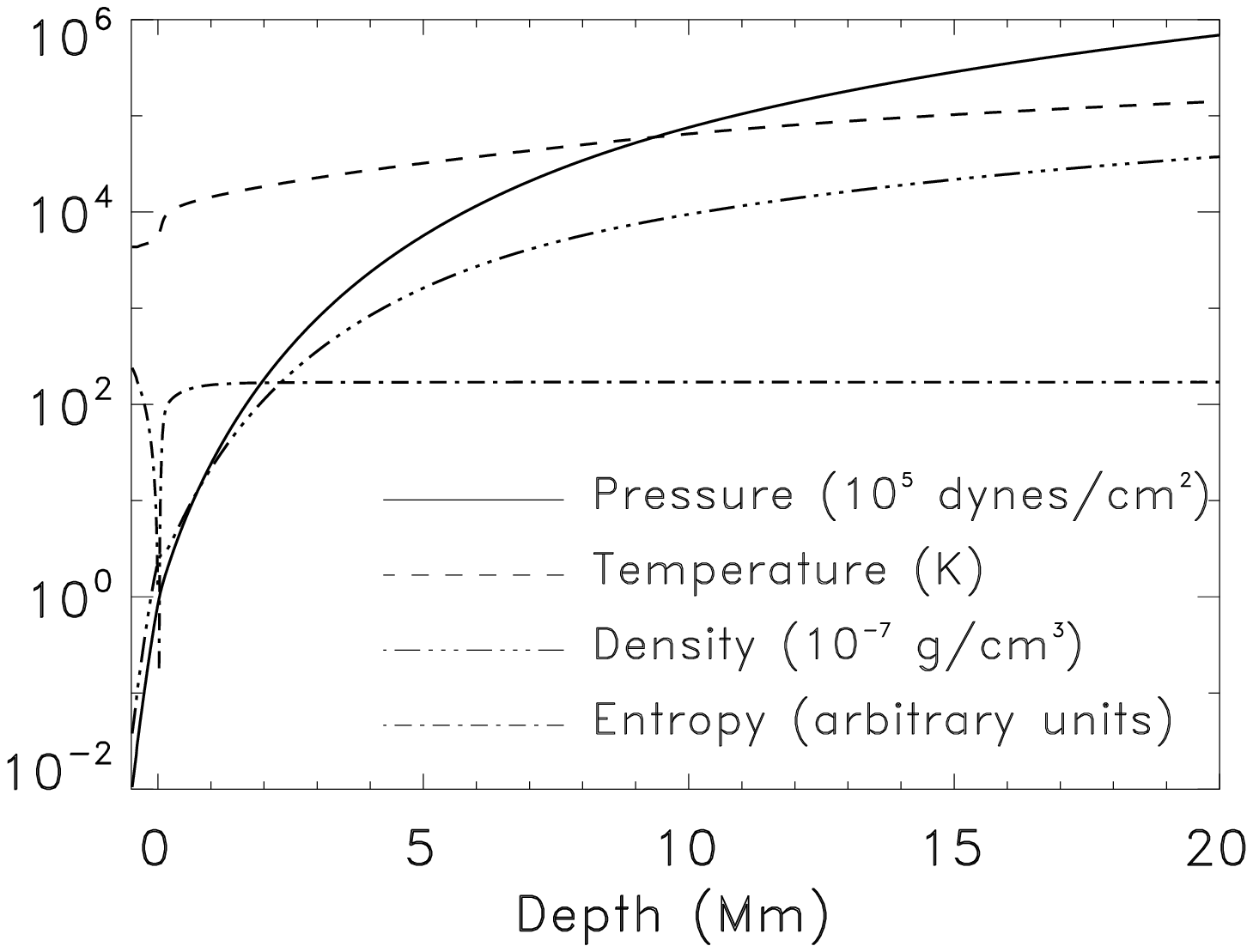}
   \includegraphics[width=0.45\textwidth]{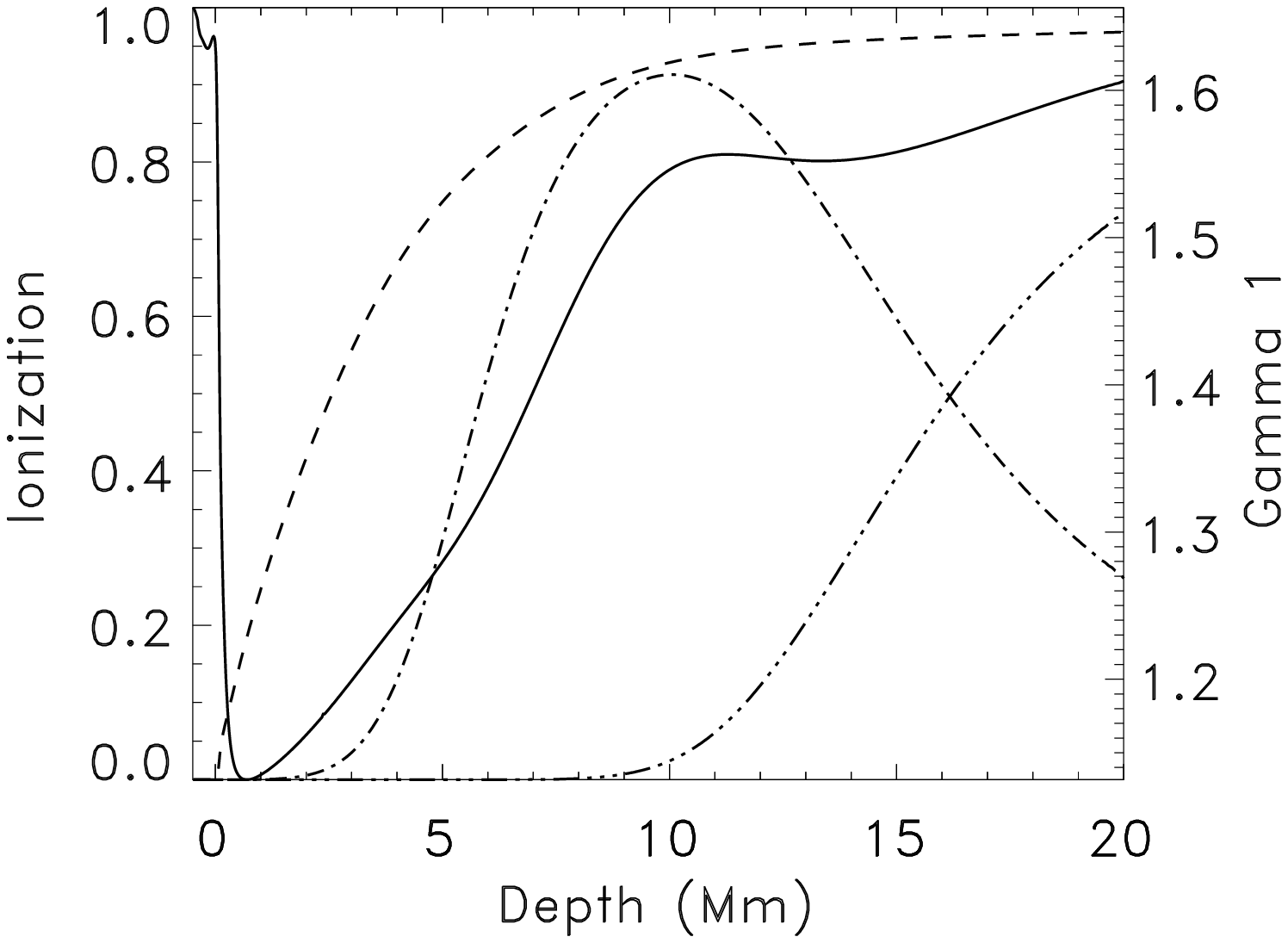}}
   \caption{Mean atmosphere of the model.  The domain includes all of
   the hydrogen and heliumI ionization zones and most of the helium II
   ionization zone (H dashed, HeI dash-dot, HeII dash-dot-dot-dot, 
   $\Gamma_1$ solid).  It covers only 10\% of the geometric depth of the
   solar convection zone, but half of the pressure scale heights.
   }
   \label{fig:atmosphere}
\end{figure}

Our domain is 48 Mm wide and from the temperature minimum to a depth of
20 Mm (Fig.~\ref{fig:atmosphere}).  We solve the equations for mass,
momentum and internal energy in conservative form plus the induction
equation for the magnetic field, for fully compressible flow, in three
dimensions, on a staggered mesh.  The code uses finite differences,
with 6th order derivative operators and 5th order interpolation
operators.  The grid is uniform in horizontal directions with resolution 
95.24 km and non-uniform in the vertical (stratified) direction with 
resolution of 12 km near the surface increasing to 75 km at the bottom 
of the domain.  Transformation
between uniform index space, where the derivatives are evaluated, and
the non-uniform physical grid is achieved by multiplying by the
Jacobian of the transformation.  The staggered mesh increases the order
of the derivative operators for the same size stencil, but the physical
variables are not aligned in space and need to be interpolated using
fifth-order interpolation operators.  Time integration is by a 3rd
order low memory Runge-Kutta scheme \cite{Kennedy99}.  Parallelization
is achieved with MPI, communicating the three overlap zones that are
needed in the 6th and 5th order derivative and interpolation stencils.
Because of the staggered mesh, div(B) is conserved to machine
precision.  However, numerical errors in div(B) that slowly accumulate
stochastically must be cleaned at intervals.  Ghost zones
are loaded at the top and bottom boundaries to permit the use of the
same spatial derivative scheme at the boundaries as in the interior.
Horizontal boundary conditions are periodic, while top and bottom
boundary conditions transmit the convective flows but partially
reflect waves.  Inflows at the bottom boundary have uniform pressure,
specified entropy and damped horizontal velocities.  Outflow boundary
values are obtained by extrapolation.  The magnetic field is made
to tend toward a potential field at the top.  That is, the ghost
zones are loaded with a potential field calculated from the values
of the vertical field just inside the boundary.  At the bottom the
magnetic field is given a specified value in inflows and extrapolated
in outflows.  The electric field in the ghost zones is calculated
from the magnetic field and the velocities (which are taken to be
constant at their values on the boundary at the top and have zero
vertical derivative at the bottom) in the ghost zones.  The magnetic
field is updated from its time derivative, which is the curl of the
electric field.
The code is stabilized by diffusion in the momentum, energy and
induction equations.

Ionization energy accounts for 2/3 of the energy transported near the
solar surface and must be included to obtain the observed solar
velocities and temperature fluctuations \cite{Stein98}.
We use a tabular equation of state, that includes local thermodynamic
equilibrium (LTE) ionization of the abundant elements as well as
hydrogen molecule formation, to obtain the pressure and temperature as
a function of log density and internal energy per unit mass.

Radiation from the solar surface produces the low entropy, high
density fluid whose buoyancy work drives the convective motions and it
also provides us with our only information on what is occurring there.
Since the optical depth is of order unity in these regions, neither
the diffusion nor optically thin approximations are valid.  We
calculate the radiative heat/cooling by solving the radiation transfer
equation in both continua and lines using the Feautrier method
\cite{Feautrier64}, assuming Local Thermodynamic Equilibrium (LTE).
The number of wavelengths for which the transfer equation is solved is
drastically reduced by using a multi-group method whereby the opacity
at each wavelength is placed into one of four bins according to its
magnitude and the source function is binned the same way
\cite{Nordlund82,Stein03,Vogler04}.  The conversion of energy
transport from convection to radiation occurs in a very thin thermal
boundary layer, which must be resolved (grid size less than 15 km) in
order to obtain the correct entropy jump and the solar convective
adiabat.  This radiation solution is validated by comparing a few
snapshots with a detailed solution having a large number of wavelength
points and many more angular quadrature points.  Calculated spectral
line widths, Doppler shifts, and asymmetries from the simulations are
in excellent agreement with observations \cite{Asplund00}.

At supergranule and larger scales, the coriolis force of the solar
rotation begins to influence the plasma motions, so $f$-plane rotation
is included in the simulation.  Angular momentum conservation
produces a surface shear layer with the surface (top of the domain)
rotating slower than the bottom of the domain, as observed in the
Sun.

We do not know the properties of the magnetic field at the depth
of 20 Mm.  We have investigated the behavior of untwisted, horizontal
flux with the same entropy as the unmagnetized ascending plasma,
advected into the computational domain by upflows from a depth of
20 Mm.  Wherever there was inflow at the bottom, uniform horizontal
field in the x-direction is advected into the computational domain.
We have calculated three cases, with field strength at the bottom
of 10, 20 and 40 kG.  In the outflows the field values are extrapolated.
At the boundaries between in- and out- flows, the field magnitude
is smoothly varied.  The boundary conditions are applied to the
electric field from which the time derivative of the magnetic field
is calculated.  The initial state in all three cases was the same
snapshot of field free hydrodynamic convection.

\section{Flux Emergence}\label{FluxEmergence} 

\begin{figure}    %%%%%%%%%%%%%%%%%% FIGURE 2 
%  \centerline{\includegraphics[width=4.8in]{I_By_h10_t33-40-48.pdf}}
   \centerline{\includegraphics[width=4.8in]{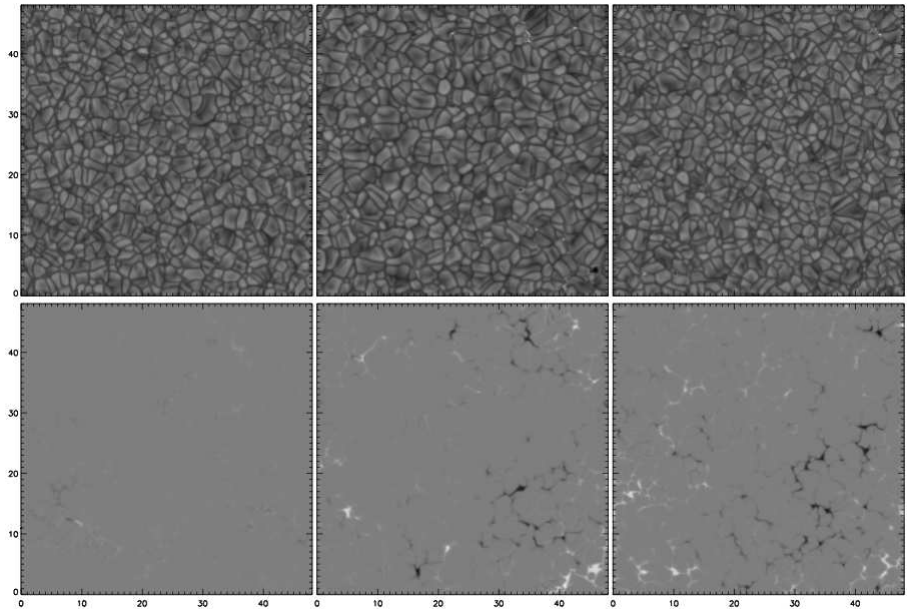}}
   \caption{Vertical field (bottom row) and emergent continuum intensity (top row) at times
   33, 40, and 48 hours after the 10 kG field started to be introduced at the bottom.  The 
   range in the field strength is [-2.8,2.84] kG.
   }
   \label{fig:fluxemergence10kG}
\end{figure}

\begin{figure}    %%%%%%%%%%%%%%%%%% FIGURE 3 
%  \centerline{\includegraphics[width=4.8in]{I_By_h20_t19-20-23.pdf}}
   \centerline{\includegraphics[width=4.8in]{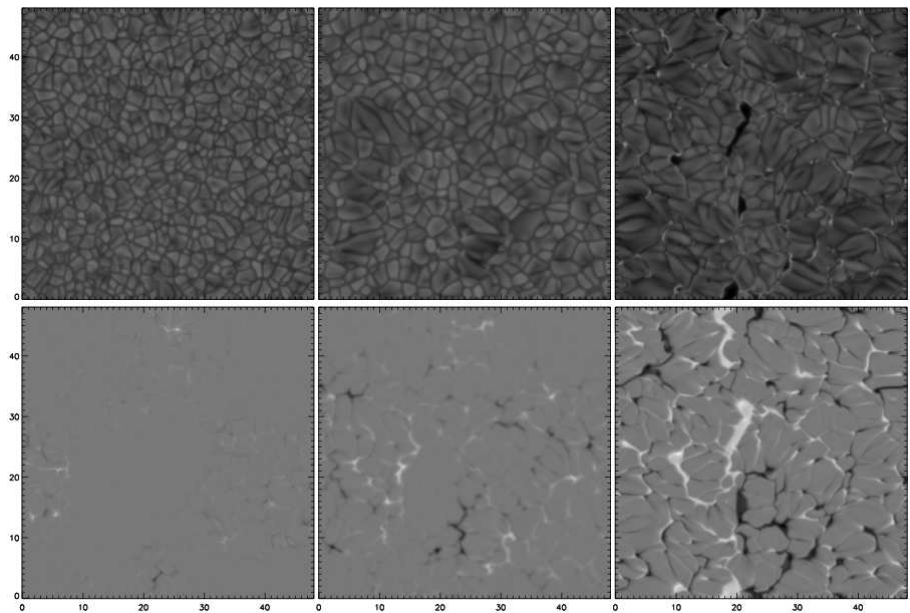}}
   \caption{Vertical field (bottom row) and emergent continuum intensity (top row) at times
   19, 20 and 23 hours after the 20 kG field started to be introduced at the bottom.  The 
   range in the field strength is [-3,3.4] kG.
   }
   \label{fig:fluxemergence20kG}
\end{figure}

Upflows carry the field toward the surface, and for the case of the
stronger fields, magnetic buoyancy increases the speed of ascent.
The typical time for fluid to rise from a depth of 20 Mm to the surface
is 32 hours.  The field in the stronger, 20 kG, case reaches the
surface in about 20 hours, assisted by magnetic buoyancy.  The field in
the weaker, 10 kG, case reaches the surface in the fluid travel time of 32
hours.  
Simultaneously, convective downflows pull portions of the field
lines downward.  Turbulence twists and stretches the fieldlines.
As the magnetic flux approaches the surface it spreads out.  In the
weaker (10 kG) case, the downflows are able to confine the field
to only a portion of the 48 Mm wide domain, but in the stronger (20
and 40 kG) cases the field spreads to over the entire domain at the
surface.  It is quickly swept by the diverging upflows to the
boundaries of the granules and more slowly to the boundaries of
meso-granule scale structures (Fig.~\ref{fig:fluxemergence10kG}).
The emerging magnetic field distorts the granules, elongating them
in the direction of the horizontal field.  As the large loop in the
20 kG case reaches the surface, it produces a large bipolar region
(Fig.~\ref{fig:fluxemergence20kG}) with a typical field strength of
2 kG at continuum optical depth unity.  The simulations have not
yet run long enough to see if the supergranule scale diverging flows
will produce a magnetic network.

\begin{figure}    %%%%%%%%%%%%%%%%%% FIGURE 4 
%  \centerline{\includegraphics[width=2.5\textwidth]{Uy_Bxz_20h_t712_y8.pdf}
%  \includegraphics[width=2.5\textwidth]{Uy_Bxz_20h_t712_y4.pdf}}
   \centerline{\includegraphics[width=2.5in]{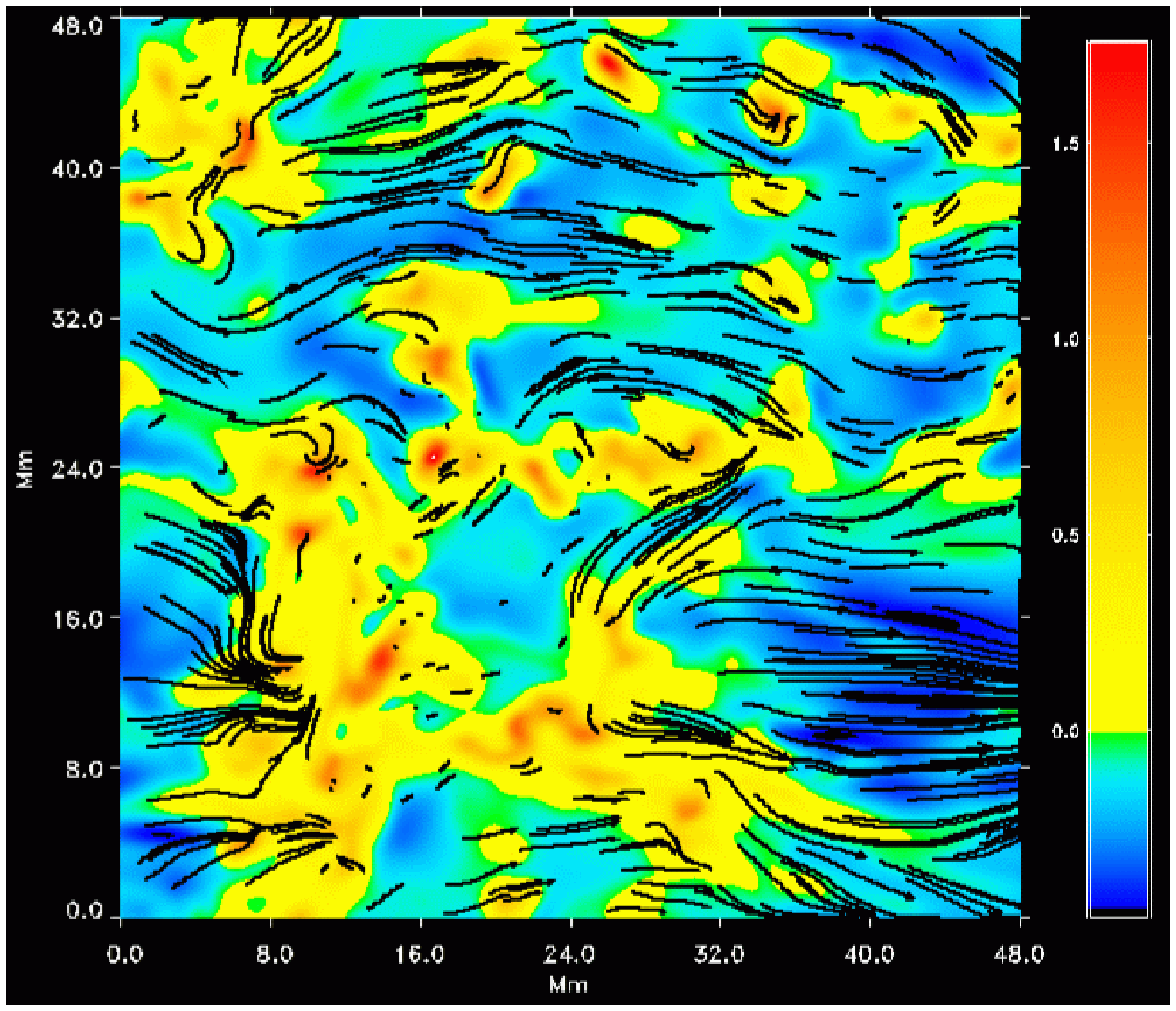}
   \includegraphics[width=2.5in]{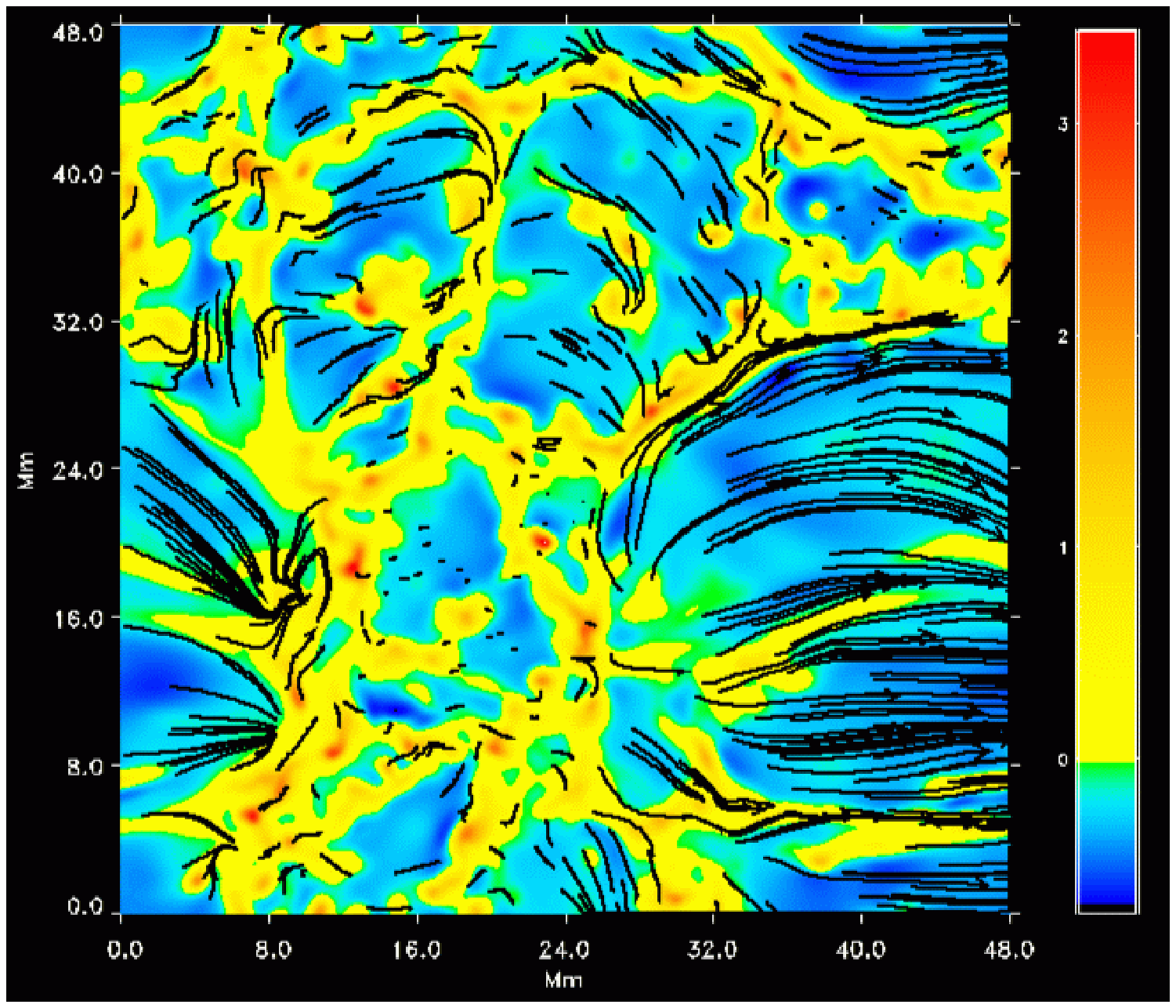}}
   \caption{Image of vertical velocity (red and yellow downward,
   blue and green upward) and horizontal magnetic field during the
   initial rise through the non-magnetic atmosphere for the 20 kG
   case at depth 8 Mm and 4 Mm.  At these intermediate depths the
   magnetic field is being advected in the upflows.  The downflows
   are still confining the magnetic flux to deeper layers and so
   are mostly field free at these depths.  The field is slowly
   spreading horizontally with decreasing depth.
   }
   \label{fig:UyBxz_20kG_t712_4_8}
\end{figure}
\begin{figure}    %%%%%%%%%%%%%%%%%% FIGURE 5 
%  \centerline{\includegraphics[width=2.5\textwidth]{Uy_Bxz_20h_t712_y2.pdf}
%  \includegraphics[width=2.5\textwidth]{Uy_Bxz_20h_t712_y1.pdf}}
   \centerline{\includegraphics[width=2.5in]{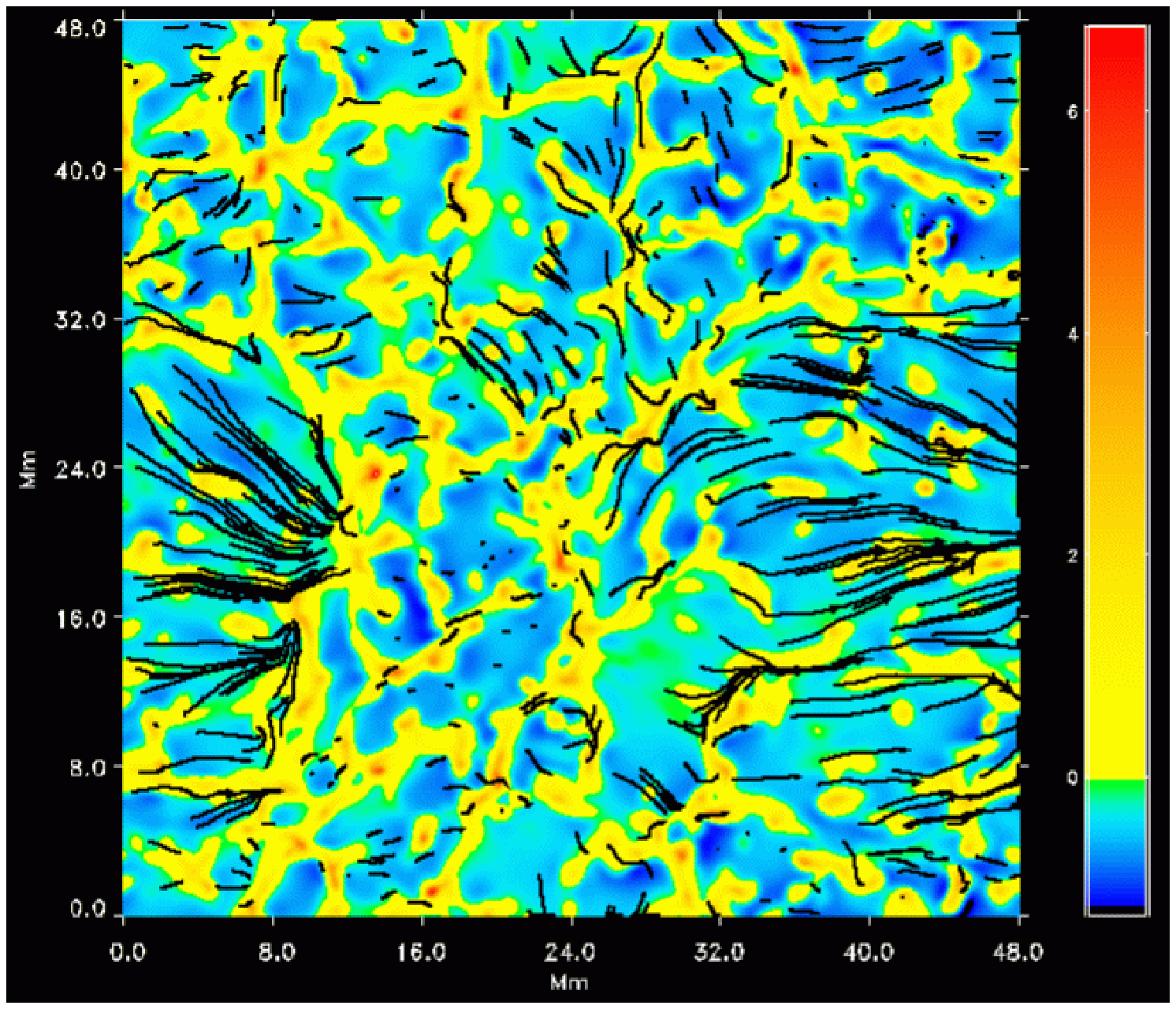}
   \includegraphics[width=2.5in]{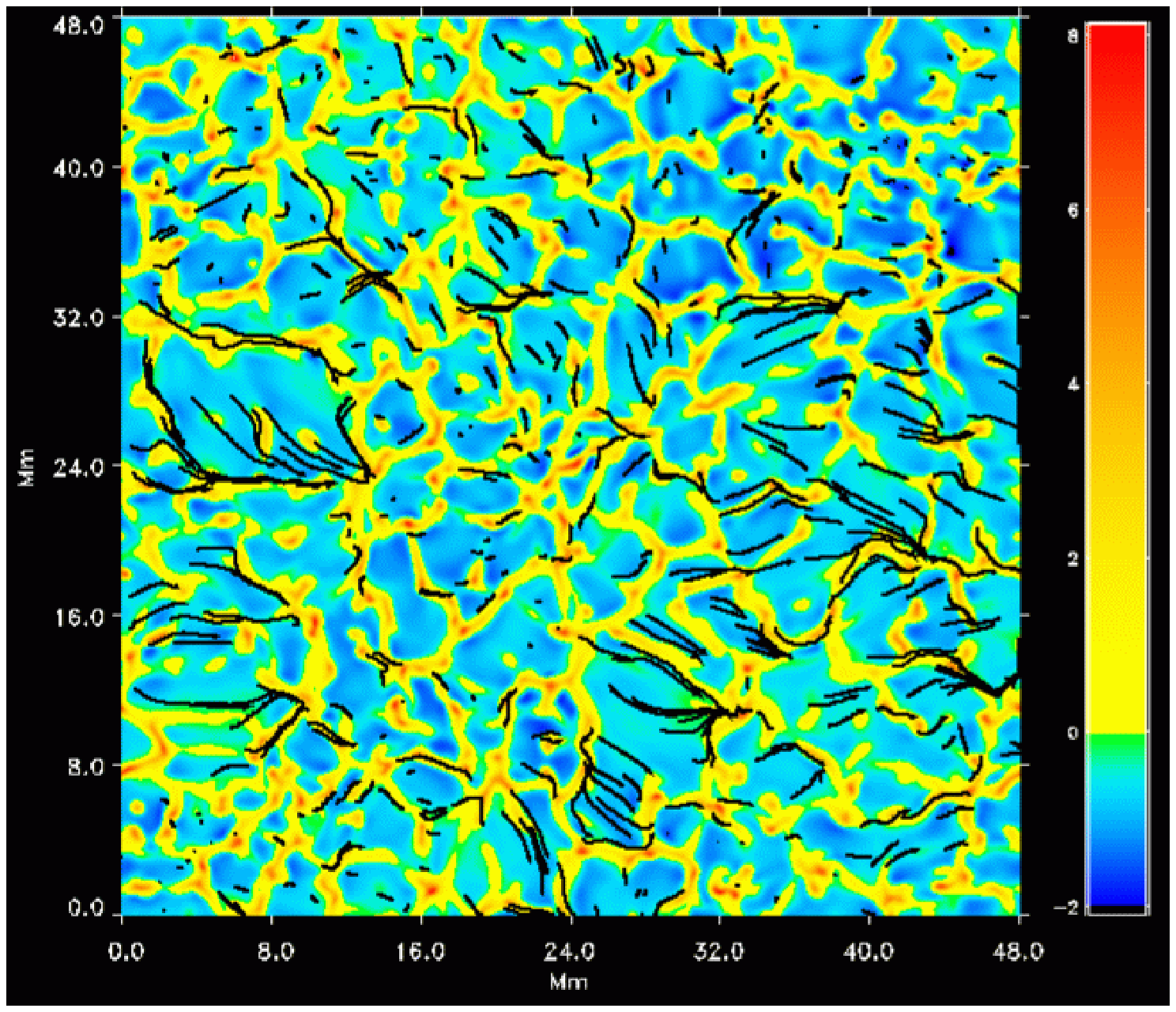}}
   \caption{Image of vertical velocity (red and yellow downward,
   blue and green upward) and horizontal magnetic field during the
   initial rise through the non-magnetic atmosphere for the 20 kG
   case at depth 2 Mm and 1 Mm.  Nearer the surface the magnetic
   field has spread farther and now almost surrounds the area of
   the large downflow at deeper levels.
   }
   \label{fig:UyBxz_20kG_t712_2_1}
\end{figure}

At the bottom of the domain, the horizontal magnetic field is entering
the domain in the supergranule scale upflows.  These upflows, plus
magnetic buoyancy, cause the field to rise to the surface.  The
downflows at these early times keep the field pinned down near the
bottom of the domain.  As a result, the downflows are nearly field
free and the magnetic field, even though untwisted, is confined
to the upflows (Fig.~\ref{fig:UyBxz_20kG_t712_4_8}).  As the surface
is approached,  the vertical velocity becomes broken up into smaller
and smaller cells and the magnetic field is spread around.  However,
the area above the large deep downflow still remains mostly field
free.  It becomes ringed with field lines diverging from it on one
side and converging toward it on the other
(Fig.~\ref{fig:UyBxz_20kG_t712_2_1}).  This is the manifestation
of the very large loop structure that is produced (Fig.~\ref{fig:lgloop}).
While the weaker field is rising, the large scale downflow
pattern at depth is changing and the magnetic field does not have such
coherent large scale loops.

\section{Magnetic Topology}\label{MagneticTopology} 

\begin{figure}    %%%%%%%%%%%%%%%%%% FIGURE 6 
%   \centerline{\includegraphics[width=0.8\textwidth]{Vvert_hslices_z3.pdf}}
    \centerline{\includegraphics[width=0.8\textwidth]{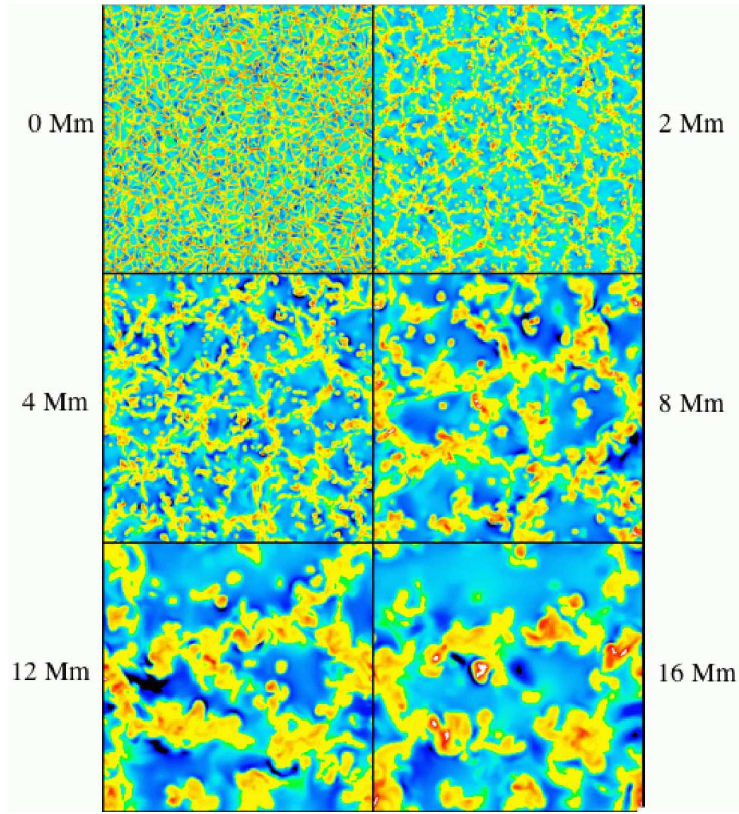}}
    \caption{Vertical velocity's horizontal pattern at the surface
    and depths 2, 4, 8, 12, and 16 Mm below the surface.  Blue and
    green are upflows.  Red and yellow are downflows.  The rms velocities are
    respectively 1.47, 0.93, 0.55, 0.35, 0.31 and 0.24 km/s.  The size of
    the horizontal cellular patterns of the vertical velocity
    increases with increasing depth as the scale height increases.
    }
   \label{fig:vvert}
\end{figure}

\begin{figure}    %%%%%%%%%%%%%%%%%% FIGURE 7 
%   \centerline{\includegraphics[width=0.9\textwidth]{B_Uy_h10_t1717_loopmix_v2.pdf}}
    \centerline{\includegraphics[width=0.9\textwidth]{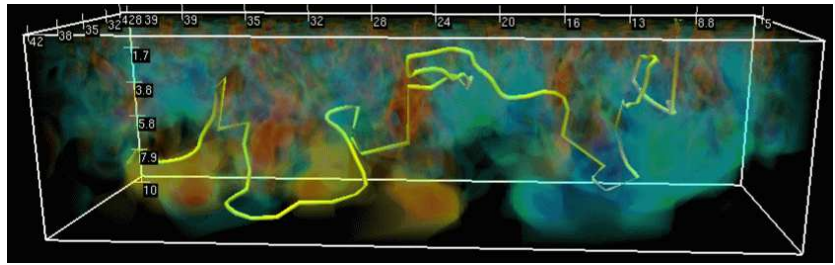}}
    \caption{Example magnetic field line for the 10 kG case, showing
    loop like structure on many scales, from 2 to 20 Mm.  Volume
    visualization is scaled vertical velocity (to make its visibility 
    uniform with depth).  Blue and green upwards,
    yellow and red downwards.  Distance scales are in Mm.
    }
   \label{fig:fieldline_loopmix}
\end{figure}

The magnetic field is transported toward the surface in the upflows
(assisted by magnetic buoyancy where the field strength is large
enough) and pushed down in the downdrafts which creates $\Omega$
and $U$ loops from the initially horizontal field.  Since the size
of the horizontal cellular structure of convection flows decreases 
as the surface is approached (from supergranule size near the
bottom to granule size near the surface, Fig.~\ref{fig:vvert})
this produces loop structures over this range of scales as well
(Fig.~\ref{fig:fieldline_loopmix}).
 
\begin{figure}    %%%%%%%%%%%%%%%%%% FIGURE 8 
%   \centerline{\includegraphics[width=0.9\textwidth]{B_Uy_h20_t796_lgloop.pdf}}
    \centerline{\includegraphics[width=0.9\textwidth]{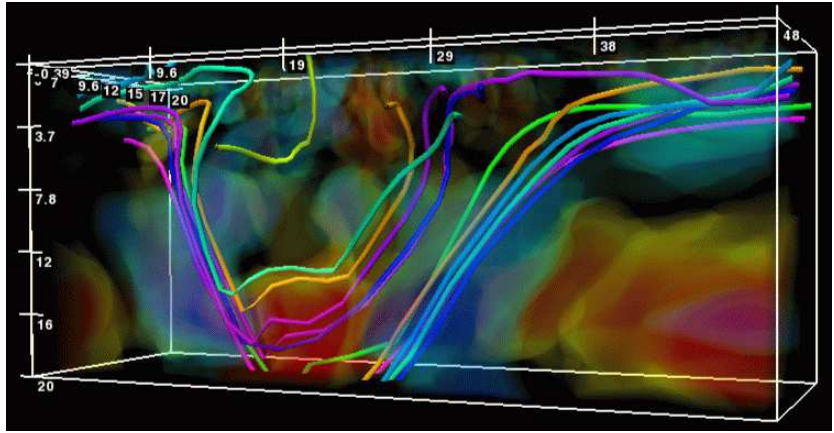}}
    \caption{Large scale magnetic loop extending the full domain
    in the x-direction (direction of the horizontal field entering
    the domain from below) and from the bottom at 20 Mm to almost
    the surface.  The volume rendering is the scaled vertical
    velocity (to make the flow at depth as visible as at the
    surface).  Red and yellow are downflows and green and blue are
    upflows.  The axis labels are megameters.
    }
   \label{fig:lgloop}
\end{figure}
\begin{figure}    %%%%%%%%%%%%%%%%%% FIGURE 9 
%   \centerline{\includegraphics[width=0.9\textwidth]{B_Uysurf_10h_t1717_v1.pdf}}
    \centerline{\includegraphics[width=0.9\textwidth]{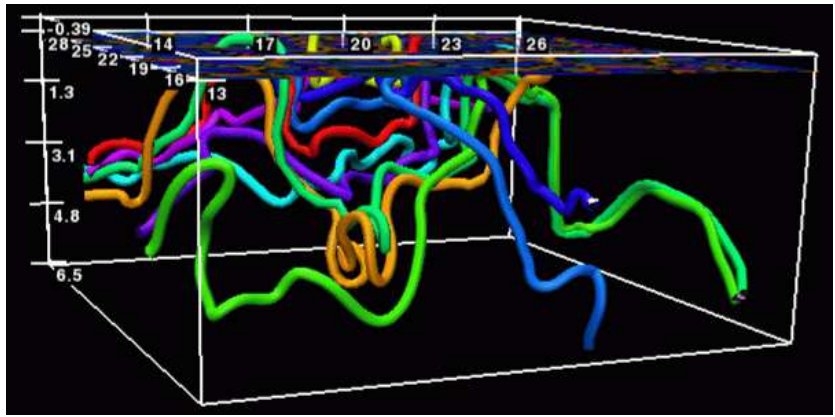}}
    \caption{Small magnetic loops near the surface.  The plane near
    the top is the visible surface with the vertical velocity
    pattern.  Note that the field lines (yellow, green, red and
    blue) that are looping over a granule interior connect very
    differently beneath the surface.
    }
   \label{fig:smloop}
\end{figure}

An example of a large scale loop structure spanning most of the
48 Mm width and 
20 Mm depth is shown in Fig.~\ref{fig:lgloop}.  The magnetic field
was buoyant as well as being transported by the upflows and was pushed
down in the downflows.  An example of small loops near the surface
is shown in Fig.~\ref{fig:smloop}.  The field lines close to the
surface are very twisted on the scale of a few megameters.  Note
that the several field lines that loop over a granule interior 
connect very differently beneath the surface because of the turbulent 
convective motions.

\begin{figure}    %%%%%%%%%%%%%%%%%% FIGURE 10
%   \centerline{\includegraphics[width=0.99\textwidth]{int_by1_pores_h20_t941.pdf}}
    \centerline{\includegraphics[width=0.99\textwidth]{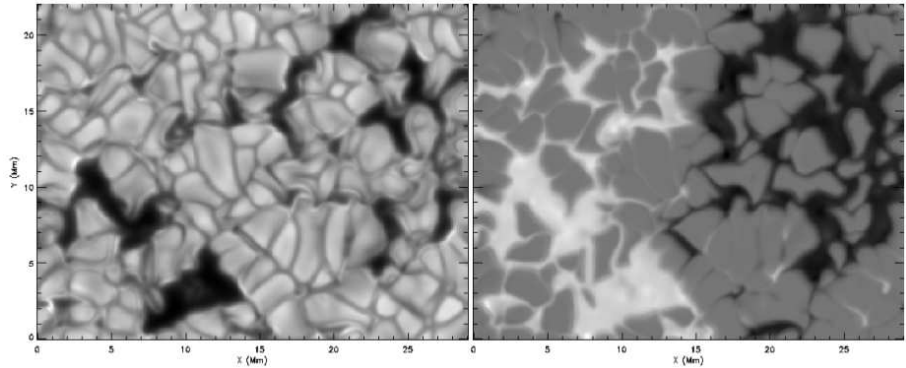}}
    \caption{Emergent continuum intensity (left) and vertical
    magnetic field at continuum optical depth unity (right) in a plage region of
    the 20 kG case.
    }
   \label{fig:pore1}
\end{figure}

In the 20 kG case, there is enough magnetic flux that it becomes
concentrated in large regions to field strengths of 2 kG at continuum optical depth
unity and produces pores (Fig.~\ref{fig:pore1}).  The intensity in
the pores is 1/4 of the average intensity.
 
\begin{figure}    %%%%%%%%%%%%%%%%%% FIGURE 11
%  \centerline{\includegraphics[width=0.49\textwidth]{AR_h20_t970_v1.pdf}
%  \includegraphics[width=0.48\textwidth,clip=]{AR_h20_t970_v2.pdf}}
   \centerline{\includegraphics[width=0.49\textwidth]{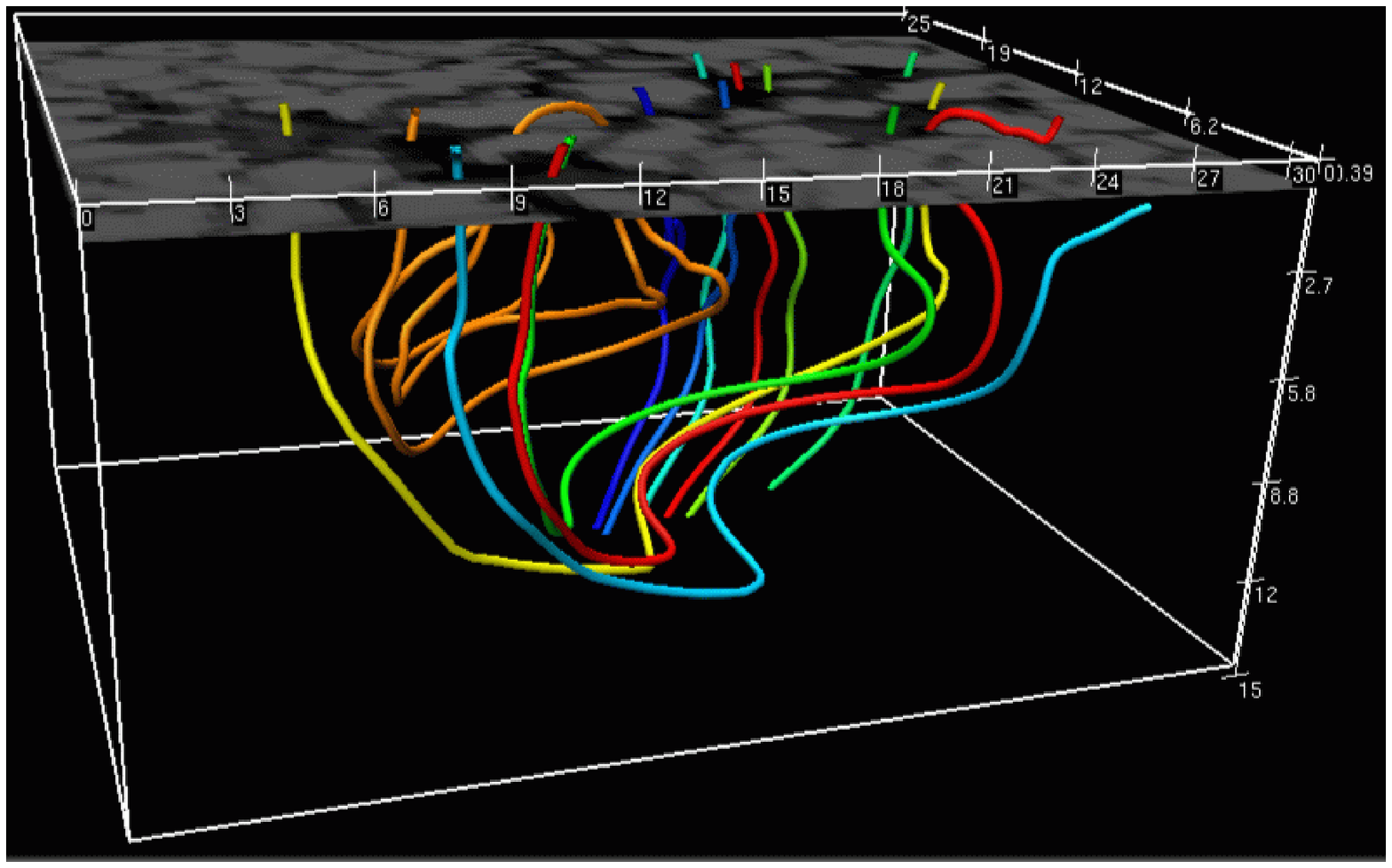}
   \includegraphics[width=0.48\textwidth,clip=]{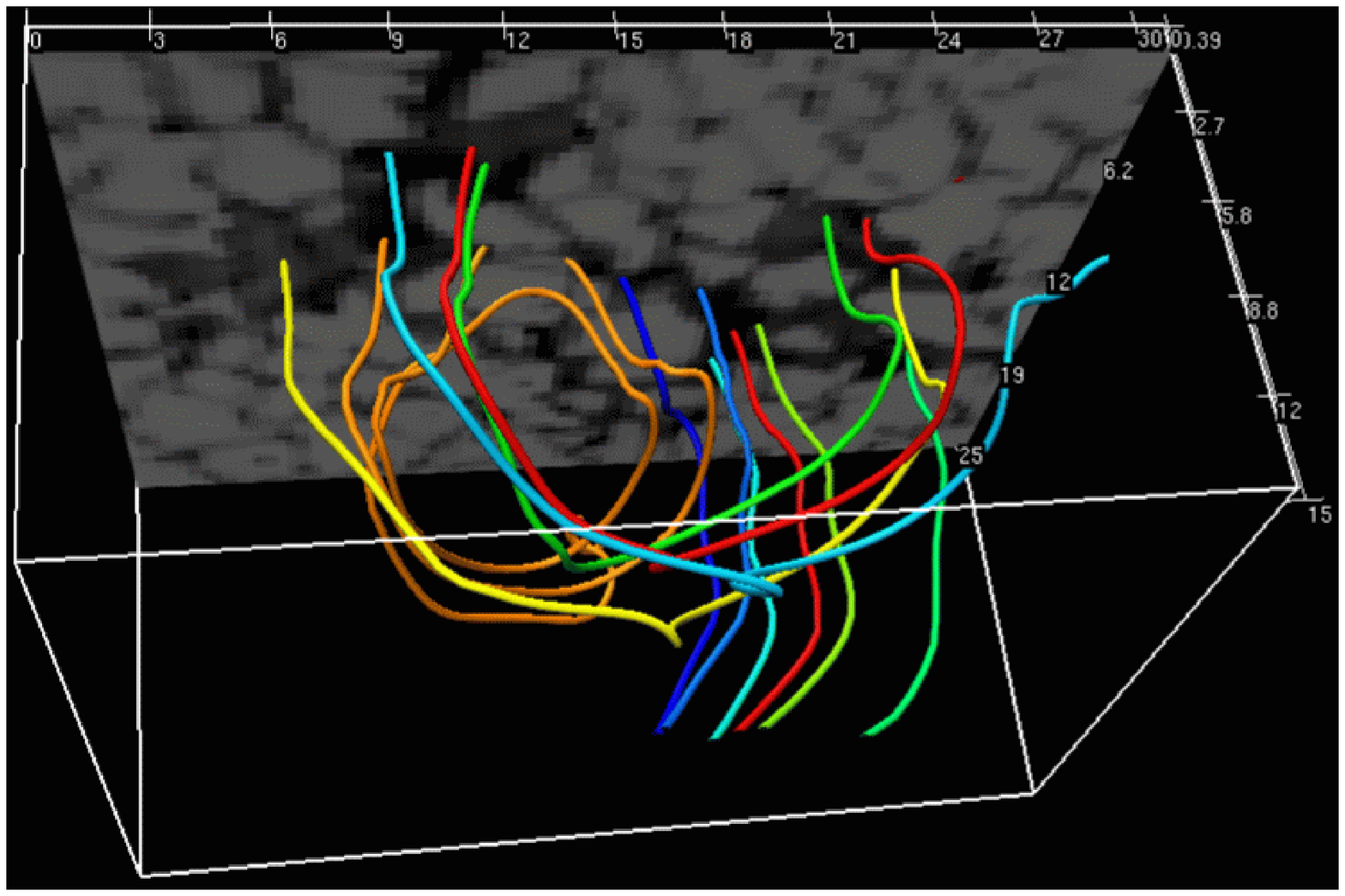}}
   \caption{Magnetic Field lines and emergent intensity in bipolar
   pore region looked at from above and below.  In this case many
   field lines are connected in a U-configuration.
   }
   \label{fig:pore2}
   \end{figure}

The magnetic field lines in this pore region are generally nearly
vertical at the surface and have a U-shape below the surface
(Fig.~\ref{fig:pore2}).  The opposite polarity field lines at the
surface generally connect at large depths (greater than 10 Mm) below
the surface.

In all three cases the magnetic energy density is larger than the
fluid kinetic energy density below 10 Mm.  However, this is not
necessarily an argument against such strong fields, since they could
be confined because they are in pressure equilibrium with their
surroundings.  These preliminary results, however, suggest that the
magnetic field can not be as strong as 20 kG at 20 Mm unless it has
a lower entropy then the field free plasma, because magnetic bubbles
develop, due to the large magnetic buoyancy, and produce large,
hot, bright granules at the surface, which are not seen.  More cases
need to be calculated and compared with observations to clarify the
properties of the magnetic field at 20 Mm depth.

\section{Application to Helioseismology}\label{Application} 

\begin{figure}    %%%%%%%%%%%%%%%%%% FIGURE 12
%   \centerline{\includegraphics[width=0.8\textwidth]{p-waves-48-zhao.pdf}}
    \centerline{\includegraphics[width=0.8\textwidth]{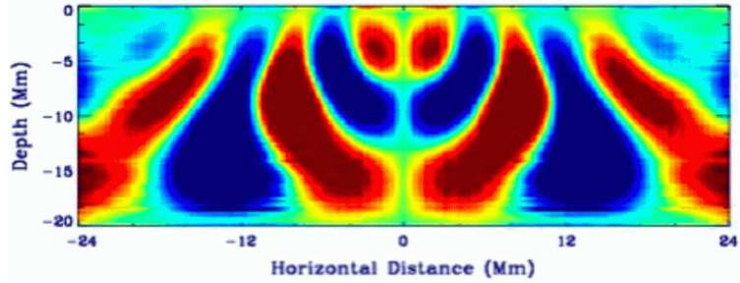}}
    \caption{Acoustic waves (visualized by the density fluctuation they produce)
    in the non-magnetic simulation propagating into the interior and refracted 
    by the increasing sound speed with depth as calculated by J. Zhao.  
    }
   \label{fig:waves}
\end{figure}

\begin{figure}    %%%%%%%%%%%%%%%%%% FIGURE 13
%   \centerline{\includegraphics[width=4.0in]{power2_nomag_mag20_cont6.pdf}}
    \centerline{\includegraphics[width=4.0in]{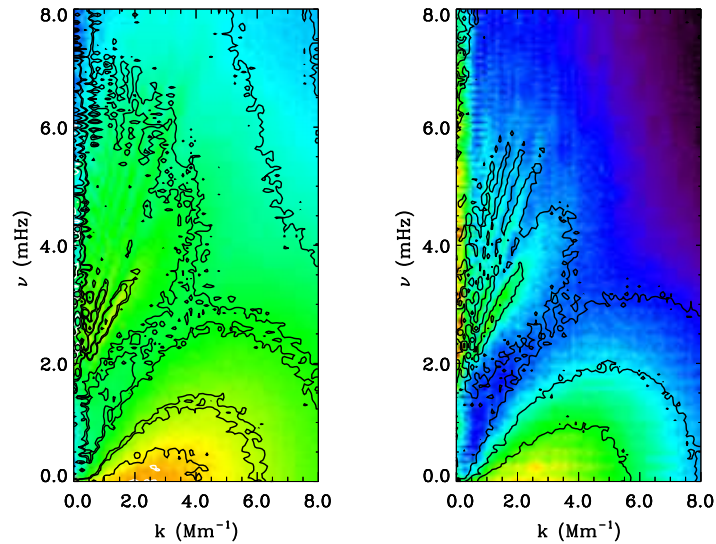}}
    \caption{$k-\omega$ diagram comparing non-magnetic and magnetic (20 kG) 
     case.  On the left the magnetic contours are overlaid on the non-magnetic
     image and vice-versa on the right.  There is no shift in the mode ridges 
     and the image on the left looks like the contours on the right and vice 
     versa, so that there does not appear to be any differences produced so 
     far in this calculation.
    }
   \label{fig:komega}
\end{figure}

$f-$ and $p-$mode waves are excited in the simulation and can be used for testing
and refining local helioseismic techniques.  Figure~\ref{fig:waves} shows an example 
of waves generated near the surface, propagating into the inerior, and refracted by
the increasing sound speed with depth.  We have made preliminary $k-\omega$ diagrams
and compared the magnetic and non-magnetic cases.  For the early stage of the flux
emergence studied so far there are no discernable differences (Fig.~\ref{fig:komega}).

\section{Discussion}\label{Discussion} 

The results we have reported here concern only the intial rise of
uniform, untwisted, horizontal magnetic flux through the near
surface, non-magnetic solar convection zone.  The main result, so
far, is that fields as strong as 20 kG at 20 Mm depth are not
possible unless they have lower entropy than the surrounding field
free plasma.  Using these simulations, we will continue to investigate
how the magnetic flux evolves over time, how the pores evolve,
whether sunspots form, whether a magnetic network is produced and
what its properties are, how flux emerges through an existing
magneto-convecting surface layer, how these simulations compare
with observations and what constraints can be placed on the properties
of the magnetic field near depths of 20 Mm.  The $f-$ and $p-$ modes
excited in these simulations will be used to investigate local
helioseismic inversion procedures.

%%%%%%%%%%%%%%%%%%%%%%%%%%%%%%%%%%%%%%%%%%%%%%%%%%%%%%%%%%%%%%%%%%%%%%%%%
  
\begin{acks}
The calculations reported here were performed with resources provided by
the NASA High-End Computing (HEC) Program through the NASA Advanced
Supercomputing (NAS) Division at Ames Research Center.  They would not
have been possible without this generous support.  Graphics were produced 
with VAPOR (www.vapor.ucar.edu) and IDL (www.ittvis.com/idl).  This work was
supported by NASA grants NNX07AO71G, NNX07AH79G and NNX08AH44G, and NSF
grant AST0605738, which is greatly appreciated.

\end{acks}

%%% BIBLIOGRAPHY %%%%%%%%%%%%%%%%%%%%%%%%%%%%%%%%%%%%%%%%%%%%%%%%%%%%%%%%%%%
\mbox{}~\\ 
   
     % format of references provided by the journal (.bst)
\bibliographystyle{spr-mp-sola}
%\bibliographystyle{spr-mp-sola-cnd} %% Alternative style: no title,
                                      % no concluding page. 

     % name your Bibtex file containing your references (.bib)
\bibliography{fluxemergence_sp09}  

     % Checking: look if the file containing the ``\bibitem'' exits
     %           so check if the .bbl file exist (bibTeX compilation)
%\IfFileExists{\jobname.bbl}{} {\typeout{}
%\typeout{****************************************************}
%\typeout{****************************************************}
%\typeout{** Please run "bibtex \jobname" to obtain} \typeout{**
%the bibliography and then re-run LaTeX} \typeout{** twice to fix
%the references !}
%\typeout{****************************************************}
%\typeout{****************************************************}
%\typeout{}}

\end{article} 

\end{document}